\documentclass{aa}  
\bibpunct{(}{)}{;}{a}{}{,} 
\usepackage{graphicx}
\usepackage{txfonts}
\usepackage{natbib}
\usepackage{lscape}
\usepackage{xcolor}
\usepackage{multicol}
\usepackage{multirow}
\usepackage{placeins}
\usepackage[colorlinks=true,allcolors=blue]{hyperref}

\newcommand{\ms}{$M_{\odot}$}
\newcommand{\rsun}{$R_{\odot}$}
\newcommand{\mzams}{$M_{\rm ZAMS}$}
\newcommand{\snii}{SN~II}
\newcommand{\sneii}{SNe~II}

\newcommand{\BV}{$(B-V)$}
\newcommand{\gr}{$(g-r)$}
\newcommand{\gi}{$(g-i)$}
\newcommand{\eq}{\,$=$\,}
\newcommand{\ml}{\,$\pm$\,}
\newcommand{\tim}{\,$\times$\,}
\newcommand{\mdot}{$\dot{M}$}\newcommand{\rcsm}{$R_{\rm CSM}$}
\newcommand{\msunyr}{$M_{\odot}$\,yr$^{-1}$}

\begin{document}

	\title{Circumstellar interaction models for the early bolometric light curve of SN~2023ixf}
	
	\author{L. Martinez \inst{1,2},
            M.~C. Bersten \inst{3,1,4},
            G. Folatelli \inst{3,1,4},
            M. Orellana \inst{5,6},
            K. Ertini \inst{3,1}
        }

	\institute{Instituto de Astrofísica de La Plata (IALP), CCT-CONICET-UNLP. 
			Paseo del Bosque S/N, B1900FWA, La Plata, Argentina\\
			\email{laureano@fcaglp.unlp.edu.ar}
   		\and
            Universidad Nacional de Río Negro. Sede Andina, Mitre 630 (8400) Bariloche, Argentina
        \and
            Facultad de Ciencias Astronómicas y Geofísicas, Universidad Nacional de La Plata, Paseo del Bosque S/N, B1900FWA, La Plata, Argentina 
        \and
            Kavli Institute for the Physics and Mathematics of the Universe (WPI), The University of Tokyo, 5-1-5 Kashiwanoha, Kashiwa, Chiba, 277-8583, Japan
        \and
            Universidad Nacional de Río Negro. Sede Andina, Laboratorio de Investigación Científica en Astronomía, Anasagasti 1463, Bariloche (8400), Argentina
        \and
            Consejo Nacional de Investigaciones Científicas y Técnicas (CONICET), Argentina 
            }

\authorrunning{L. Martinez et al.}
\titlerunning{CSM interaction models for  SN~2023ixf}

\date{Received XXX; accepted XXX}

\abstract 
{Type II supernovae (\sneii) show growing evidence of interaction with circumstellar material (CSM) surrounding their progenitor stars as a consequence of enhanced mass loss during the last years of the progenitor's life, although the exact mechanism is still unknown.
We present an analysis of the progenitor mass-loss history of SN~2023ixf, a nearby \snii\ showing signs of interaction.
First, we calculate the early-time ($<$\,19~days) bolometric light curve for SN~2023ixf based on the integration of the observed flux covering ultraviolet, optical and near-infrared bands, and black-body extrapolations for the unobserved flux.
Our calculations spot the sudden increase to maximum luminosity and temperature, in addition to the subsequent fall, displaying an evident peak.
This is the first time that this phase can be precisely estimated for a \snii\ showing interesting characteristics as: 1) slope changes during the rise to maximum luminosity; and 2) a very sharp peak with a maximum luminosity of $\sim$\,3\tim10$^{45}$~erg\,s$^{-1}$. 
We use the early-time bolometric light curve of SN~2023ixf to test the calibrations of bolometric corrections against colours from the literature. In addition, we include the observations of SN~2023ixf into some of the available calibrations to extend their use to earlier epochs.
Comparison of the observed bolometric light curve to \snii\ explosion models with CSM interaction suggests a progenitor mass-loss rate of \mdot\eq3\tim$10^{-3}$\,\msunyr confined to 12000\,\rsun\ ($\sim$\,8\tim$10^{14}$\,cm) and a wind acceleration parameter of $\beta$\eq5.
This model reproduces the early bolometric light curve, expansion velocities, and the epoch of disappearance of interacting lines in the spectra.
This model indicates that the wind was launched $\sim$\,80~yr before the explosion.
If the effect of the wind acceleration is not taken into account, the enhanced wind must have developed over the final months to years prior to the SN, which may not be consistent with the lack of outburst detection in pre-explosion images over the last $\sim$\,20~yr before explosion.
}


\maketitle
\section{Introduction}
\label{sec:intro}

Type II supernovae (\sneii\footnote{Throughout this paper we use the denomination `SNe II' to refer to hydrogen rich core-collapse supernovae excluding type IIn, IIb, and SN~1987A-like events.}) are the result of the explosion of massive stars ($\gtrsim$\,9\,\ms) that have retained a hydrogen-rich envelope at the end of their evolution. \sneii\ are characterised by prominent hydrogen lines in their spectra \citep{minkowski41,filippenko97} and are the most common type of core-collapse SNe \citep{shivvers+17}.
Direct detections of progenitors in pre-explosion images provide strong evidence for red supergiant (RSG) stars as \snii\ progenitors \citep[e.g.,][]{vandyk+12a,smartt15}.

Hydrogen-rich SNe are sub-classified based on their photometric and/or spectral characteristics. Some of these objects show prevalent narrow emission lines in their spectra and luminous light curves \citep{schlegel90,arcavi17}.
These events are referred to as type IIn SNe. The characteristics of this subgroup are attributed to the interaction of the SN ejecta with a pre-existing dense circumstellar material (CSM). This CSM is the result of a high mass-loss rate during the last stage of the progenitor evolution.

A significant fraction of \emph{normal} \sneii\ also show narrow emission features, disappearing within hours to days after explosion \citep{bruch+21}, thus suggesting that the spatial extension of the CSM is small and that the progenitor experienced enhanced mass loss shortly before core collapse \citep[e.g.,][]{yaron+17}.
The SN shock wave breaks out from the progenitor surface emitting high-energy photons that excite and ionise the CSM; moreover, the continuous interaction between the shock wave and the CSM converts kinetic energy into radiation that also ionises the material outwards the shock front. 
The narrow emission features are a consequence of the recombination of the slow-expanding ionised CSM \citep{khazov+16,dessart+17,smith17}.

SN~2023ixf is a \snii\ \citep{perley+23_tns} discovered on 2023 May 19 17:27:15.00 UT in the galaxy M101 \citep{itagaki23_tns}.
The proximity to this object allowed several detections of the progenitor candidate in pre-explosion images taken with the \textit{Hubble Space Telescope}, the \textit{Spitzer Space Telescope} and ground-based telescopes. 
The analysis of these images results in a variable RSG candidate obscured by dust, whose luminosity is consistent with the evolution of a star with an initial mass of \mzams\eq10$-$15\,\ms\ \citep{szalai+23_atel,pledger+23,kilpatrick+23,neustadt+23,jencson+23,xiang+23}.
At the same time, the analysis of the progenitor variability implies an initially more massive star of \mzams\eq20\ml4\,\ms\ \citep{soraisam+23}.
Additionally, the study of the stellar populations in the vicinity of the site of explosion of SN~2023ixf infer a progenitor initial mass of \mzams\eq17$-$19\,\ms\ \citep{niu+23}.
The progenitor could not be detected in X-rays and ultraviolet (UV) pre-explosion images \citep{kong+23_atel,matsunaga+23_atel,basu+23_atel}.
Pre-explosion observations disfavour the presence of any outbursts in the last $\sim$20~yr \citep{neustadt+23,jencson+23,dong+23}.

After the discovery, intensive photometric, spectroscopic, and polarimetric follow up have been carried out \citep[e.g.,][]{hosseinzadeh+23,grefenstette+23,teja+23,vasylev+23}.
Early-time spectra show narrow emission features during the first week after discovery, which indicate the presence of a dense CSM \citep{sutaria+23_atel,yamanaka+23,teja+23,jacobson-galan+23,bostroem+23,smith+23}.

In the present paper, we attempt to estimate the physical properties of the CSM surrounding the progenitor of SN~2023ixf by modelling the early-time bolometric light curve and evolution of the expansion velocity.
We note that the characteristics of the wind producing the CSM has to be consistent not only with the aforementioned observables of SN~2023ixf but also with the epoch of disappearance of the narrow emission features and the absence of outbursts ---at least--- during the last $\sim$20~yr before explosion.

There are only a small number of objects observed as early and intensively as SN~2023ixf; therefore, it is a great opportunity to calculate and analyse the early bolometric light curve of a \snii. 
The early follow-up of SN~2023ixf allowed to calculate the bolometric light curve before the maximum luminosity, during the rise to peak.
This is of particular interest given that this has only been observed in a small number of previously discovered \sneii.
Given the exceptional temporal and wavelength coverage of SN~2023ixf observations, the analysis of this early phase can provide important information about the shock wave emergence.
In addition, the calculation of the early-time bolometric light curve allows us to estimate bolometric corrections (BCs) and to extend the calibrations of BC against optical colours previously established in the literature \citep{martinez+22a} to earlier epochs.
In a forthcoming paper, we derive the progenitor and explosion physical properties of SN~2023ixf by modelling its complete bolometric light curve and photospheric velocity evolution (Bersten et al. in prep.).

In this work, we adopt a Cepheid-based distance of 6.85\ml0.15\,Mpc \citep{riess+22}.
For the explosion epoch, there are various constraints thanks to the large number of non-detections close to the discovery date. We adopt MJD\,60082.75 as the explosion date, following the analysis of \citet{hosseinzadeh+23} and the non-detections by \citet{mao+23_tns}.
The Milky Way reddening in the direction of SN~2023ixf is $E(B-V)_{\rm MW}$\eq0.008\,mag \citep{schlafly+11}, while we adopt a host-galaxy reddening of $E(B-V)_{\rm host}$\eq0.031\,mag based on the equivalent widths of \ion{Na}{I} lines (\citealt{lundquist+23_tns}, see also \citealt{smith+23}). We assume a Galactic extinction law from \citet{cardelli+89} with $R_{V}$\eq3.1.

The present paper is organised as follows. 
Section~\ref{sec:blc} describes the methodology to calculate the bolometric light curve of SN~2023ixf.
Section~\ref{sec:bc} inspects the currently available calibrations for BCs versus optical colours and presents an extension of the calibrations previously found by \citet{martinez+22a} by including SN~2023ixf in the analysis.
Section~\ref{sec:modelling} presents the modelling to the early-time bolometric light curve of SN~2023ixf and the derived physical properties of the CSM.
In Sect.~\ref{sec:discussion}, we discuss the scenario that produces the CSM and compare with the results from the literature.
We provide our concluding remarks in Sect.~\ref{sec:conclusions}.

\section{Bolometric light curve}
\label{sec:blc}

\begin{figure}
\centering
\includegraphics[width=0.49\textwidth]{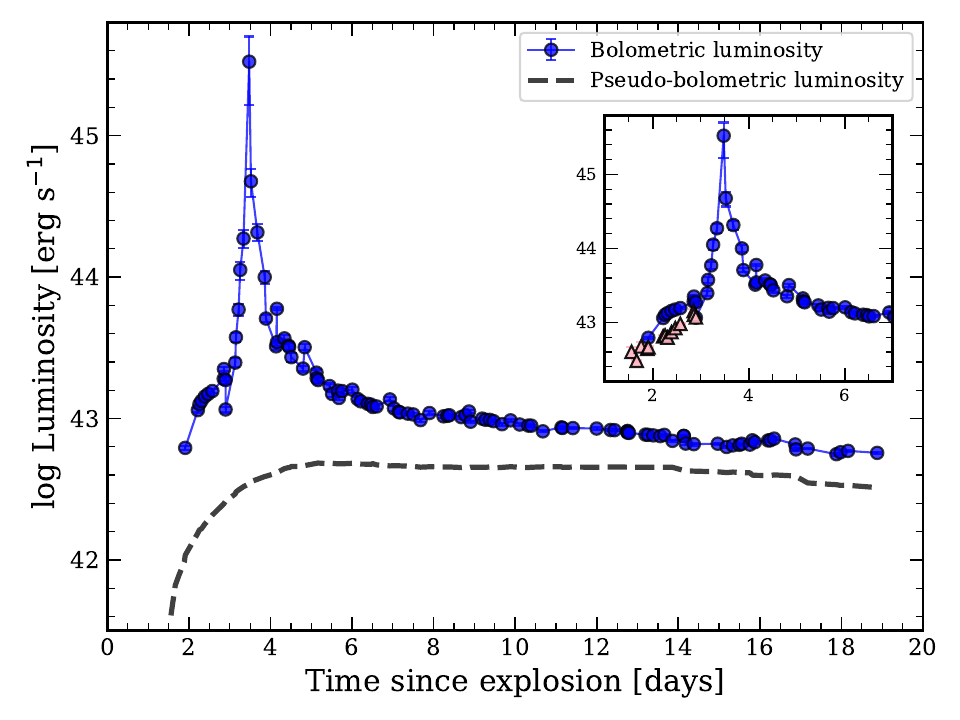}
\caption{Early bolometric light curve for SN~2023ixf (blue dots). The dashed line represents the pseudo-bolometric light curve. The inset plot shows the first week of evolution of the bolometric luminosity. In this plot, the pink triangles are the bolometric luminosities when observed UV data are not taken into account in the calculation method. In most cases the error bars are smaller than the dot size.}
\label{fig:sn2023ixf_bol}
\end{figure}

\begin{figure}
\centering
\includegraphics[width=0.49\textwidth]{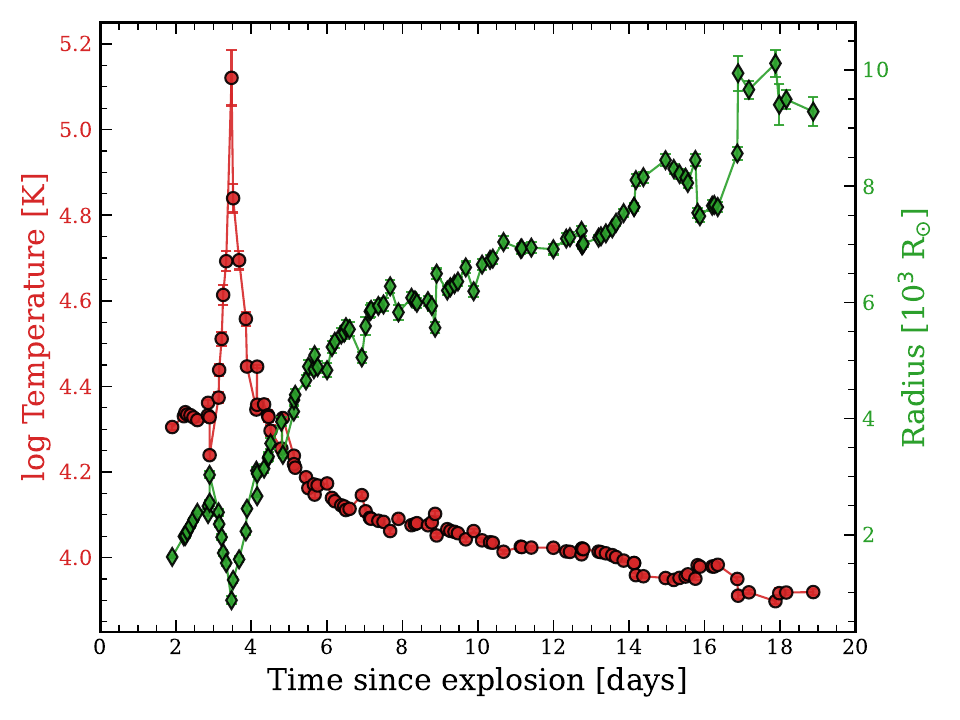}
\caption{Evolution in time of black-body fit parameters for SN~2023ixf: temperature (red dots) and radius (green diamonds).}
\label{fig:bb_pars}
\end{figure}

The main goal of this work is to derive physical properties for the mass-loss history of the progenitor of SN~2023ixf near core collapse, based on comparing models with early-time observations (<\,19\,days). 
The models are computed using a 1D code that simulates the explosion of the SN and calculates bolometric luminosities, among other observables \citep[see Sect.~\ref{sec:modelling} for additional details]{bersten+11}.
Therefore, in a first stage, we estimate bolometric luminosities for SN~2023ixf.


SN~2023ixf has been monitored since shortly after its discovery with an exceptional cadence and wavelength coverage.
In order to estimate a reliable bolometric light curve, we collect publicly-available multi-band photometric data from the literature and additional photometry reported through The Astronomer's Telegrams\footnote{\url{https://www.astronomerstelegram.org/}} and TNS Astronotes\footnote{\url{https://www.wis-tns.org/astronotes}} services.
Specifically, we gathered UV, optical ($UBVu'g'r'i'z'$ filters) and near-infrared (NIR, $JHK_{s}$ filters) magnitudes from \citet{teja+23}, and optical photometry ($BVRIg'r'i'$) from \citet{balam+23_tns}, \citet{davanzo+23_tns}, \citet{fowler+23_tns}, \citet{kendurkar+23_tns}, \citet{sgro+23_tns}, and \citet{vannini23a_tns,vannini23b_tns,vannini+23c_tns}.
The UV data presented in \citet{teja+23} correspond to $UVW2$, $UVM2$, and $UVW1$ filters from the Ultraviolet Optical Telescope \citep{roming+05} on board the \emph{Swift} Observatory \citep{gehrels+04}.
The entire data set covers from 0.3 to 19~days after explosion, which allow us to analyse the early SN emission and to estimate the physical properties of the CSM.

The estimation of the bolometric luminosities was performed in the same manner as in \citet{martinez+22a}. 
This method consists in the integration of the observed fluxes, which in the present study represents the spectral energy distribution (SED) of SN~2023ixf from mid-UV ---when available--- to NIR wavelengths. In addition, the calculation method assumes that the SN emits as a black body at the unobserved wavelengths (see details below).

The early-time photometry of SN~2023ixf is characterised by a high cadence of observations. However, magnitude values are not always available at a given epoch for all the observed bands, which are necessary to produce reliable black-body fits to the observed SED.
We obtained a complete set of magnitudes at each observed epoch performing \texttt{loess} non-parametric regressions using the \texttt{ALR} code\footnote{\url{https://github.com/olrodrig/ALR}} described in \citet{rodriguez+19}.
Observed $UVW2$ and $UVM2$ light curves have a small number of observations, therefore, these light curves were not interpolated. 
Extrapolations were not allowed for any band.

Having photometric measurements or interpolated magnitudes in all observed bands at each epoch of observation (with the exception of $UVW2$ and $UVM2$), we proceeded with the bolometric luminosity estimation method.
We transformed magnitudes into monochromatic fluxes at the mean wavelength of the filter using the transmission functions provided by the SVO filter service \citep{svo2012,svo2020}, taking into account that the collected data are available in different photometric systems.
The monochromatic fluxes were then integrated using the trapezoidal method and the observed flux was estimated at each epoch of observation.

To estimate the unobserved flux at shorter and longer wavelengths we assume that the SN emission in those regimes is well described by a black-body model.
At early times, this assumption is mostly correct. As the SN ejecta expands and cools, the UV emission starts to depart from a black-body model as a consequence of the increasing line blanketing produced by iron-group elements.
However, with our collected data set, we note that black-body models are still consistent with the observed UV emission at least up to $\sim$15\,days from explosion (after this epoch there are no available observations at UV bands).
Therefore, it is not necessary to remove the bluest bands from the black-body fitting as it is, in general, for later observations \citep[see e.g.][]{bersten+09,faran+18,martinez+22a}.
Black-body fits were carried out only for observational epochs with at least four observed/interpolated bands.

Once we found a black-body model that fits the observed SED, we assume that the extrapolated flux at longer wavelengths is simply the emission of the black-body model between the reddest observed band and infinity (known as the IR correction). At the same time, the extrapolated flux at shorter wavelengths is the emission of the black-body model between the bluest observed band and zero wavelength (known as the UV correction).
The sum of observed flux and the extrapolated fluxes from the black-body model equates the bolometric flux.

To take the magnitude uncertainties into account, we calculated the bolometric flux via a Monte Carlo procedure. For each of the two thousand simulations, we randomly sampled broadband magnitudes assuming a Gaussian distribution centred at the magnitude value with a standard deviation equal to the magnitude uncertainty. 
Then, the observed flux was integrated, the best-fitting black-body model was found, and the IR and UV corrections were estimated.
The mean bolometric flux of the two thousand simulations was calculated and taken as the bolometric flux. We took the standard deviation of the distribution as the uncertainty of the luminosity. This procedure was repeated at every epoch of observation. Finally, the bolometric flux was transformed into luminosity using the distance to the SN.
The bolometric light curve was calculated from 1.9 to 18.9~days after explosion. 

Figure~\ref{fig:sn2023ixf_bol} shows the resulting bolometric light curve for SN~2023ixf. 
In addition, Fig.~\ref{fig:sn2023ixf_bol} shows the pseudo-bolometric light curve for SN~2023ixf, which is defined as the integration of the observed flux in the optical and NIR regimes.
At early times, the differences between both light curves are significant. This behaviour indicates the great contribution of the UV to the bolometric flux at these epochs. Moreover, the absence of the UV flux erases the luminosity peak.
Therefore, if the photometric coverage is limited to optical and redder bands, or the unobserved flux in the UV is not taken into account, the peak in the bolometric light curve is lost.
Eventually, the differences become smaller because the SN ejecta cools and the UV emission decreases, while the SN emission in the optical increases.

The early bolometric light curve of SN~2023ixf consists of a rapid rise time of 3.47\,days to maximum at $\log$\,$L_{\rm bol}$\eq45.5$^{+0.18}_{-0.30}$ ($M_{\rm bol}$\eq$-$25.08\,$\pm$\,0.54~mag). 
This is the first time ---to our knowledge--- that such a detailed rise to maximum and sharp peak are observed in bolometric luminosities, having a large wavelength coverage to make reliable estimations of the bolometric luminosities, and using similar techniques.
At the epoch of maximum luminosity, the black-body model fits observed fluxes in the following bands: $UVOT$-$B$, $B$, $g$, $UVOT$-$V$, $V$, $r$, $i$, $z$, $J$, $H$, and $K$; resulting in a black body with a temperature of $\sim$\,1.3\tim10$^{5}$\,K.
After peak, the luminosity drops $\sim$2.3\,dex in the following 1.5\,days.
Then, the luminosity starts a slower decline, at least up to day~19 post-explosion.

During the luminosity rise, SN~2023ixf shows different slopes to reach the maximum luminosity (see Fig.~\ref{fig:sn2023ixf_bol}).
First, the luminosity increases almost linearly up to 2.2~days post-explosion. 
Then, the luminosity starts a slower rise up to day~2.9. Finally, the light curve rises up to maximum with a single slope ---much steeper than in earlier times--- from day~2.9.
This epoch matches with the last epoch of observation ---before maximum--- having \emph{Swift} data in the UV.
Specifically, before maximum luminosity, the $UVW1$-band light curve is available until 2.9~days post-explosion, while only a single data point in the $UVM2$ band was obtained.

In order to test the influence of the available UV data on the rise to maximum luminosity we calculate the early bolometric light curve for SN~2023ixf again, but this time neglecting the \emph{Swift} data in the UV regime, both from the black-body fitting procedure and the integration of the observed SED. This is shown in the inset plot of Fig.~\ref{fig:sn2023ixf_bol} as pink triangles (only the first three days are shown given that these are the epochs with most UV observations). 
This process results in a bolometric light curve without the slope changes mentioned above, and with lower luminosities before day~2.9 post-explosion.
The lower luminosities are obtained because the \emph{Swift} data in the UV (mostly in the $UVW1$ band) are more luminous than the predicted flux from black-body models at the mean wavelength of the $UVW1$ filter, when the $UVW1$ data are neglected from the calculation. 
This means that black-body fits ignoring the available UV data underestimate the UV extrapolation.
Therefore, the observed $UVW1$ data produce black-body models that peak at shorter wavelengths, i.e. hotter black-body models, causing larger UV corrections and higher temperatures during this time interval (see Fig.~\ref{fig:bb_pars}).
Although the lack of UV observations after 2.9~days introduces more uncertainty in the bolometric luminosities, this lack does not explain the sudden increase in temperature and luminosity that occurs just after that time.
In summary, the bolometric luminosities before day~2.9 strongly depends on the UV flux, which is very intense at these epochs. Unfortunately, at the moment there are no more public data in the UV.

Figure~\ref{fig:bb_pars} shows the black-body parameters obtained from the fits. 
Before the luminosity peak, the temperature shows values between $\log$\,T\,$\sim$\,4.3$-$4.4, while the radius increase by a factor of $\sim$2.
Then, the temperature suddenly increases to a value of $\log$\,T\,$\sim$\,5.1 in less than 0.5~days, coincident with the maximum luminosity. At the same time, the black-body radius takes smaller values.
After the luminosity peak, the black-body temperature (radius) decreases (increases) almost monotonically.

\section{Bolometric corrections}
\label{sec:bc}

In Sect.~\ref{sec:blc}, we estimate bolometric luminosities for SN~2023ixf through direct integration of the observed flux (covering UV, optical, and NIR bands) and assuming that the SN emits as a black body at shorter and longer ---unobserved--- wavelengths.
This is the most accurate method to estimate bolometric luminosities when extensive wavelength coverage is available.
The use of bolometric corrections to convert broadband magnitudes into bolometric magnitudes is a more frequent technique when the photometric coverage is limited only to optical filters.
From the work of \citet{bersten+09}, where the authors developed calibrations between BCs and optical colours, several other studies have analysed these relations \citep[e.g.][]{lyman+14,pejcha+15a}.
More recently, \citet{martinez+22a} presented updated calibrations of BC against optical colours using the most homogeneous and largest sample of \snii\ bolometric light curves.

The unprecedented early-time bolometric light curve of SN~2023ixf, characterised by a high cadence of observations and the wide wavelength coverage of the broadband data, allows us to examine the calibrations of bolometric corrections versus colour found in the literature (Sect.~\ref{sec:bc_test}) and to extend previous calibrations to bluer colours (i.e. to earlier times, Sect.~\ref{sec:new_bc_calibrations}).

\subsection{Testing calibrations of bolometric corrections}
\label{sec:bc_test}

\begin{figure*}
\centering
\includegraphics[width=0.49\textwidth]{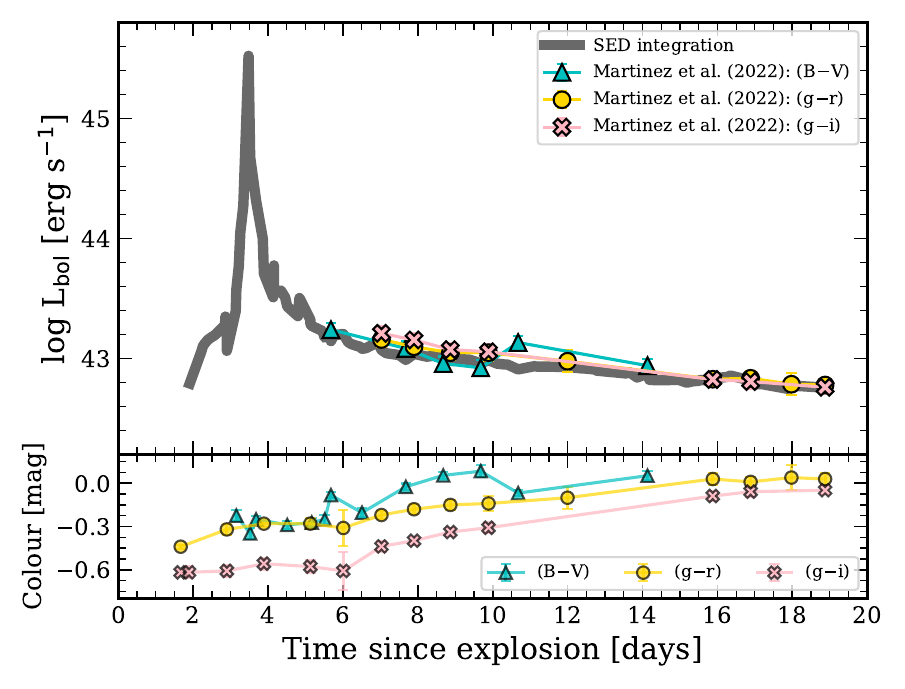}
\includegraphics[width=0.49\textwidth]{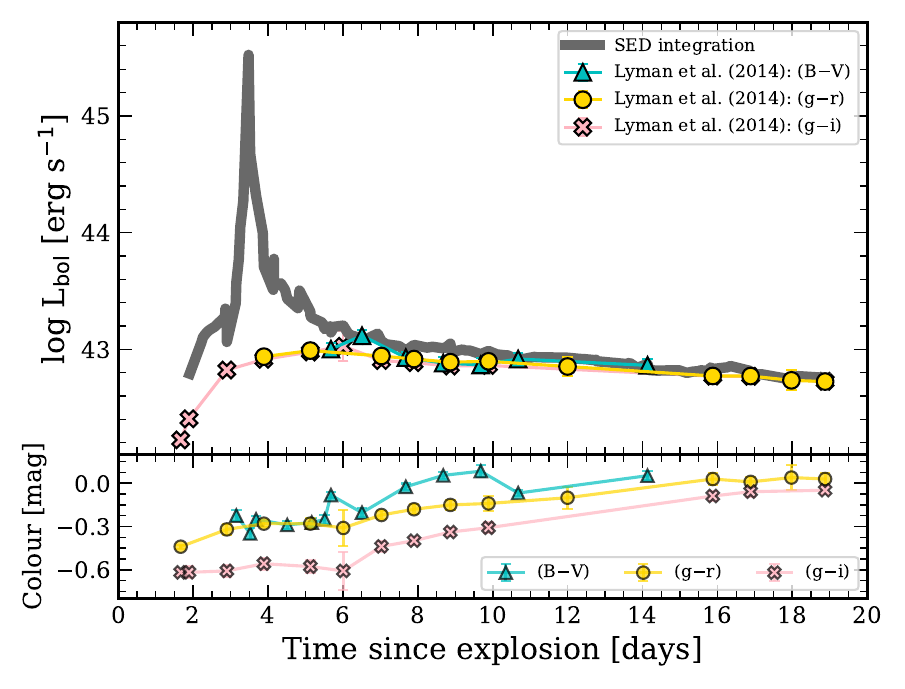}
\includegraphics[width=0.49\textwidth]{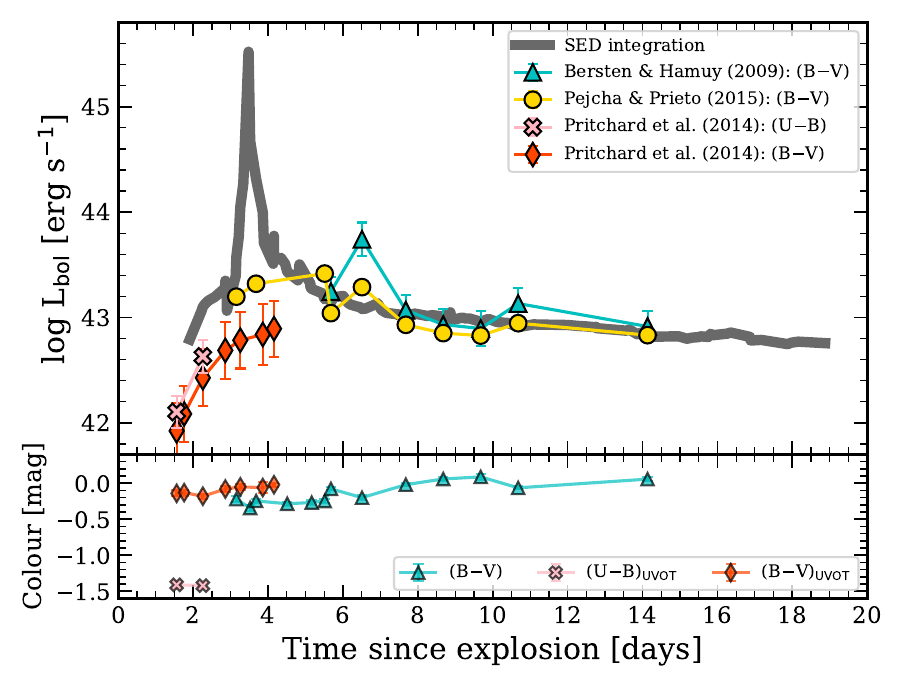}
\includegraphics[width=0.49\textwidth]{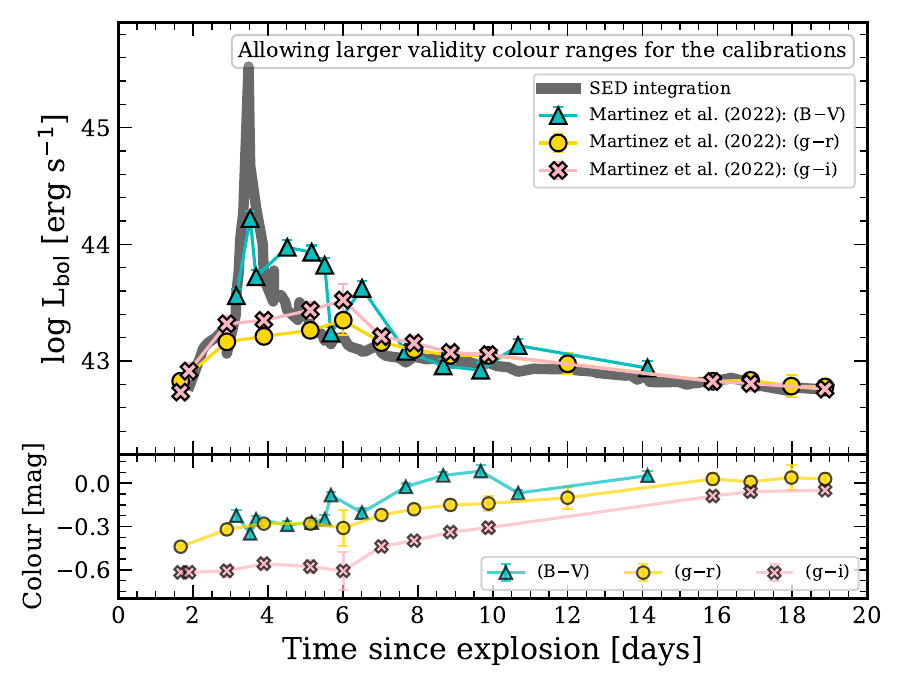}
\caption{Bolometric light curve for SN~2023ixf calculated from the integration of the observed flux plus black-body extrapolations (thick grey line, referred to as `SED integration') in comparison with those calculated from calibrations of bolometric corrections versus colours found in the literature: \citet{martinez+22a} \textit{(top-left panel)}, \citet{lyman+14} \textit{(top-right panel)}, \citet{bersten+09}, \citet{pejcha+15a}, and \citet{pritchard+14} \textit{(bottom-left panel)}. The bottom-right panel shows bolometric light curves using the calibrations by \citet{martinez+22a} when larger validity ranges of colours are allowed.}
\label{fig:sn2023ixf_bc}
\end{figure*}

In this section, we compare the bolometric light curve of SN~2023ixf estimated in Sect.~\ref{sec:blc} with those constructed employing the calibrations of bolometric corrections from the literature.
Specifically, we compare with the calibrations from \citet{bersten+09}, \citet{lyman+14}, \citet{pritchard+14}, \citet{pejcha+15a}, and \citet{martinez+22a}. 
Figure~\ref{fig:sn2023ixf_bc} shows the results of the analysis. Each panel also shows the colour curves used for the calculation. The details of the comparison are found below.

\citet{martinez+22a} presented calibrations of BCs versus \BV, \gr, and \gi\ colours, with the latter two colour indices showing the smallest dispersions. These BC calibrations are distinguished according to the phase in which the SN is found. For the comparison, we utilised the calibrations that corresponds to the `cooling phase', since these are the most appropriate for our data set. 
In addition, these calibrations were performed with photometric data points in the natural system of the Swope telescope at Las Campanas Observatory \citep{contreras+10}. Therefore, the first step is to convert our data into the corresponding photometric system.
Vega magnitudes were transformed into AB system using the conversion values published in \citet{blanton+07}.
We then used the magnitude offsets from \citet{krisciunas+17} to convert AB magnitudes into the natural system of the Swope telescope. At that moment, the calibrations of BCs were applied.

We find good agreements between the bolometric light curve calculated in Sect.~\ref{sec:blc} (referred to as `SED integration' in Fig.~\ref{fig:sn2023ixf_bc}), and those calculated using the BC calibrations from \citet{martinez+22a} (Fig.~\ref{fig:sn2023ixf_bc}, top-left panel). The bolometric light curve constructed with the BC calibration against \gr\ produces the most similar light curve to that observed. At the same time, the predicted bolometric luminosities using the calibration versus \gi\ are slightly brighter than those estimated using the \gr\ colour index.
The bolometric light curve using the BC calibration against \BV\ agrees well between days~6 and 10. After day~10 the predicted bolometric luminosities overestimate the observations.
We note that earlier estimations are not possible because the colour values are bluer than the validity ranges of the calibrations (see below for predicted bolometric luminosities if the validity colour ranges are not considered).

The top-right panel of Fig.~\ref{fig:sn2023ixf_bc} compares our bolometric light curve and those predicted using the BC calibrations for the cooling phase from \citet{lyman+14}. These authors constructed calibrations for several colour indices. However, we show comparisons only to calibrations using \BV, \gr, and \gi\ given that the other colour indices present a small number of data points.
The bolometric light curve computed using BC calibrations versus \BV\ shows good agreement with our, with the exception of the data points between 7 and 10~days post-explosion. At those epochs, the predicted luminosities underestimate our estimation by $\sim$0.1\,dex.
The predicted bolometric light curves using \gr\ and \gi\ colours show similar behaviours. In both cases, the luminosity is underestimated, especially during the first 5~days post-explosion.
For the \gi\ colour, the rise to maximum is much smoother than the calculated with our procedure, similar to the behaviour of the pseudo-bolometric light curve (see Fig.~\ref{fig:sn2023ixf_bol}).

The bottom-left panel of Fig.~\ref{fig:sn2023ixf_bc} shows a comparison with several other calibrations found in the literature.
We chose to compare with the BC calibrations versus \BV\ from \citet{bersten+09} and \citet{pejcha+15a}. The other BC calibrations from these latter two papers cannot be well compared due to the small number of data points for the colours involved [($B-I$) and ($V-I$) in the case of \citet{bersten+09} and ($B-R$) and ($B-I$) for \citet{pejcha+15a}].
The bolometric light curve calculated with the BC calibration from \citet{bersten+09} present two data points  ---around days~6.5 and 10.5 post-explosion--- much brighter than those using the SED integration method. With the exception of these values, the luminosity agree well with our estimate.
The BC calibration from \cite{pejcha+15a} agrees well with our bolometric light curve at some epochs. However, other epochs show a variable behaviour. This behaviour can possibly be explained due to the irregular conduct of the \BV\ colour curve, which could also explain the over-luminous data points in the comparison with the BC calibration from \citet{bersten+09}.
Additionally, we compare to the BC calibrations from \citet{pritchard+14}. These calibrations were performed for ($U-B$) and ($B-V$) colours using \emph{Swift}+UVOT filters; therefore, we use the available \emph{Swift}+UVOT photometry for this comparison. For both BC calibrations, the resulting bolometric light curves are much dimmer than that estimated via SED integration.

Finally, we used the BC calibrations from \citet{martinez+22a} again, but this time without considering the validity ranges of colours. This means that we extrapolated the calibrations to bluer colours.
The predicted bolometric light curves using the extrapolated BC calibrations versus \gr\ and \gi\ shows remarkable good agreement at these early epochs, with the exception of the value around day~6. However, we note the large error bars in the \gr\ and \gi\ colour curves at that epoch, arising predominantly from the $g$-band magnitude.
Surprisingly, the BC calibration versus \BV\ predicts the behaviour of the rise to maximum luminosity and the subsequent drop, although the following data points clearly overestimate the luminosity from the SED integration method.
However, as stated before, we note the variable behaviour of the \BV\ colour curve.
This analysis shows that the BC calibrations from \citet{martinez+22a} are a satisfactory method to estimate bolometric luminosities, particularly the calibrations versus \gr\ and \gi\ colours.

\subsection{Calibrations of bolometric corrections including SN~2023ixf}
\label{sec:new_bc_calibrations}

\begin{table}
\caption{Coefficients of the polynomial fits to the bolometric corrections versus optical colours.}             
\label{table:bc_calibration}
\centering          
\resizebox{0.49\textwidth}{!}{
    \begin{tabular}{lccccccc}
    \hline\hline\noalign{\smallskip}
    Colour & Range & $c_{0}$ & $c_{1}$ & $c_{2}$ & $c_{3}$ & $c_{4}$ & $\sigma$ \\
    \hline\noalign{\smallskip}
    ($g-r$) & ($-$0.43, 1.09) & $-$0.353 & 1.643 & $-$3.574 & 1.474 & --- & 0.133 \\
    \noalign{\smallskip}
    ($g-i$) & ($-$0.60, 1.15) & $-$0.220 & 0.738 & $-$2.137 & 0.913 & --- & 0.125 \\
    \noalign{\smallskip}
    ($B-V$) & ($-$0.35, 1.16) & $-$0.704 & 4.013 & $-$7.985 & 6.904 & $-$2.357 & 0.206 \\
    \hline
    \end{tabular}}
\tablefoot{BC = $\sum_{k=0}^{n} c_{k}(colour)^{k}$, where colour is taken from the first column. The last column ($\sigma$) represents the standard deviation about the fit.}
\end{table}

\begin{figure}
\centering
\includegraphics[width=0.49\textwidth]{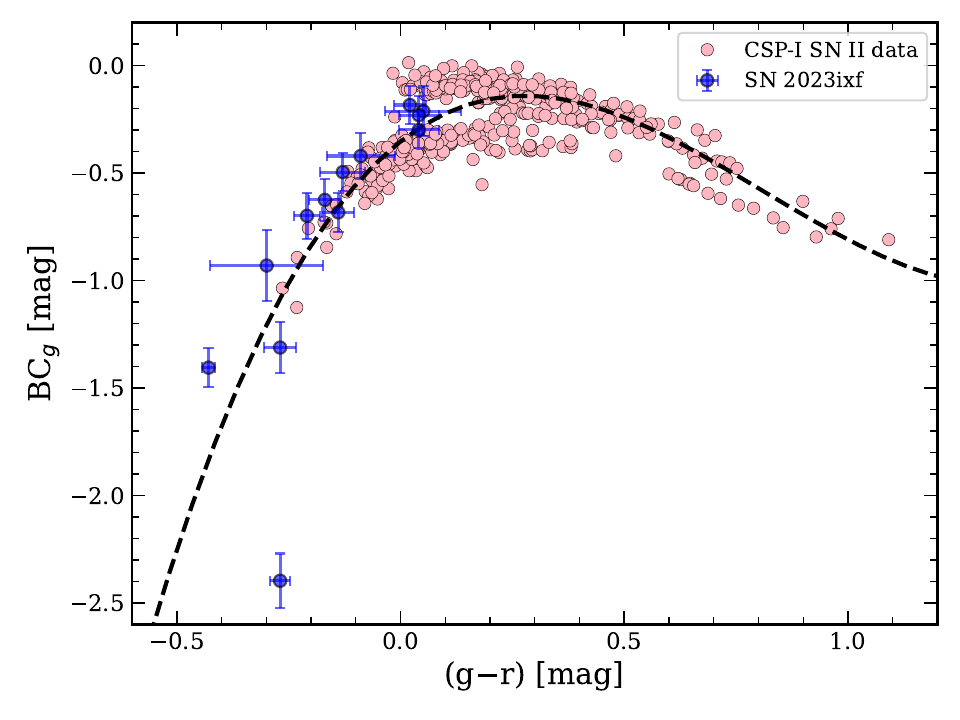}
\includegraphics[width=0.49\textwidth]{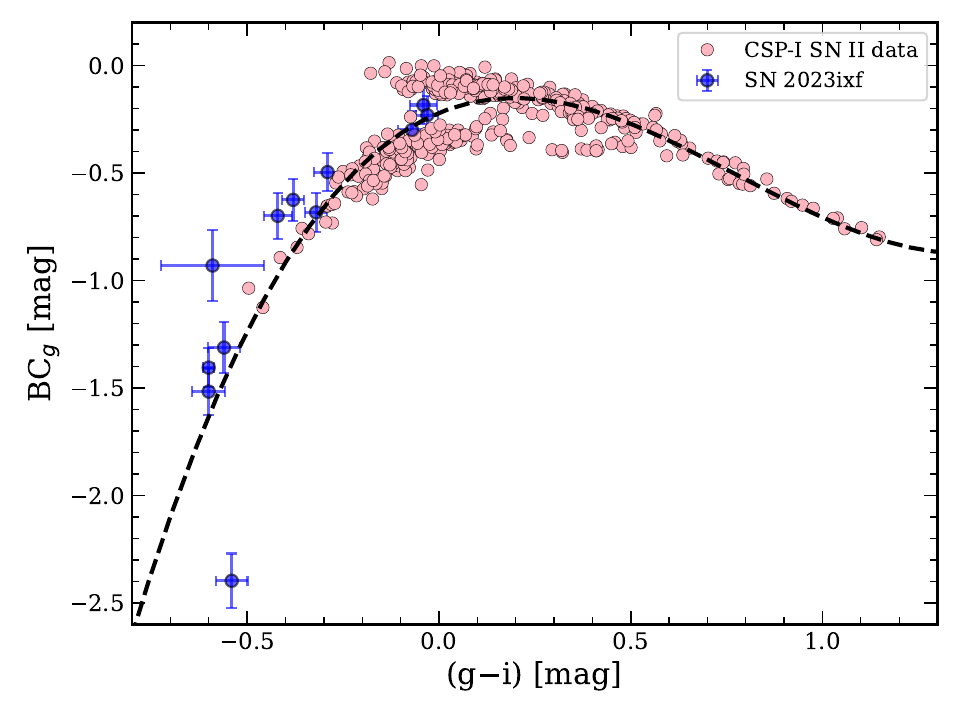}
\includegraphics[width=0.49\textwidth]{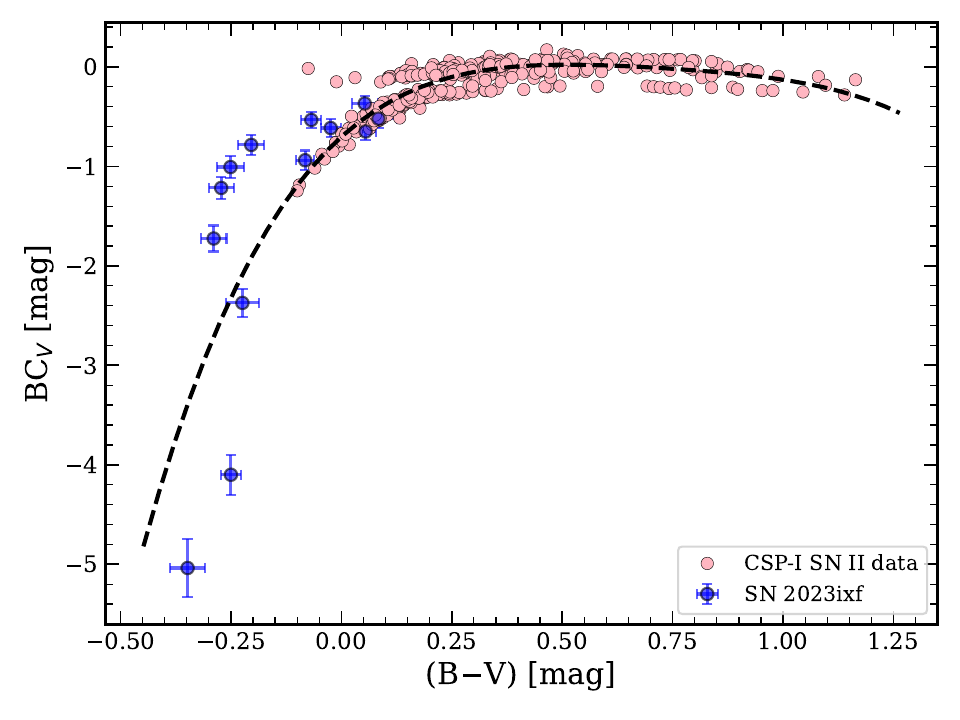}
\caption{Bolometric corrections relative to the $g$ band as a function of ($g-r$) colour \textit{(top panel)} and ($g-i$) colour \textit{(middle panel)}, and relative to the $V$ band as a function of \BV\ colour \textit{(bottom panel)}. SN~2023ixf is presented as blue dots, while pink dots represent the cooling phase of the \sneii\ in the CSP-I sample \citep[see][]{martinez+22a}. The dashed lines shows the fit to the data. The errors in the CSP-I \snii\ data are not plotted for better visualisation.}
\label{fig:bc_cal}
\end{figure}

In Sect.~\ref{sec:bc_test}, we show that the BC calibrations from \citet{martinez+22a} adequately reproduce the bolometric light curve of SN~2023ixf.
In addition, SN~2023ixf has an unique early-time bolometric light curve, due to the high observational cadence that resolves the rise to maximum luminosity, and the extensive wavelength coverage ---from UV to NIR--- that allows us to estimate precise bolometric luminosities.
This motivates us to incorporate the early-time data of SN~2023ixf to the BC calibrations from \citet{martinez+22a} in order to extend the calibrations (corresponding to the `cooling phase') to bluer colours (i.e. to earlier times).

The BC calibrations from \citet{martinez+22a} were performed using a sample of 74~\sneii\ observed by the Carnegie Supernova Project-I \citep{hamuy+06} using the facilities of the Las Campanas Observatory.
Therefore, we used the same data, in addition to those from SN~2023ixf, to construct new BC calibrations.

First, we converted the bolometric luminosities of SN~2023ixf into bolometric magnitudes. By definition
\begin{equation}
    M_{\rm bol} = M_{\odot,\rm bol} - 2.5\,\mathrm{log}_{10}\,
    \left( \frac{L_{\rm bol}}{L_{\odot,\rm bol}}  \right) \,,
\end{equation}
where $L_{\odot,\rm bol}$~=~3.845~$\times$~10$^{33}$~erg~s$^{-1}$ and $M_{\odot,\rm bol}$~=~4.74~mag are the luminosity and the absolute bolometric magnitude of the Sun \citep{allen2000}.
We then calculated the bolometric corrections for SN~2023ixf employing the definition,
BC$_{j}$\,=\, $m_{\rm bol} - m_{j}$, where $m_{j}$ is the extinction-corrected magnitude in the band $j$ of the SN, and $m_{\rm bol}$ is its the bolometric magnitude. 
Finally, we looked for calibrations between the bolometric corrections and the same three colour indices.

Figure~\ref{fig:bc_cal} displays the bolometric correction relative to the $g$~band (BC$_{g}$) as a function of \gr\ and \gi\ colours (top and middle panels, respectively).
Figure~\ref{fig:bc_cal} also includes polynomial fits to the data computed via Markov chain Monte Carlo (MCMC) methods using the \texttt{python} package \texttt{emcee} \citep{emcee}.
We used third order polynomial fits for the calibrations comprising \gr\ and \gi\ colours.
We find good agreement between the polynomial fits and the data, except for the lowest BC$_{g}$ value in both plots (BC$_{g}$\eq$-$2.40\,mag). This value corresponds to the epoch when the luminosity peak is taking place.

We also searched for calibrations between the bolometric correction relative to the $V$ band (BC$_{V}$) as a function of \BV. This is shown in the bottom panel of Fig.~\ref{fig:bc_cal}. 
For this case, we utilised a fourth order polynomial to fit the early-time data of SN~2023ixf.
We did not find any improvement in the BC calibration versus \BV\ colour with respect to that obtained using the CSP-I \snii\ data, i.e. towards \BV\ values lower than $-$0.10~mag.
The lowest two BC$_{V}$ values (BC$_{V}$\eq$-$5.04 and $-$4.10~mag) corresponds to the peak time.
However, we note that these calibrations should be considered more uncertain for the bluest colours for the following reasons: 1) we are using only one \snii\ at these colour ranges; and 2) the steep dependence of the BC with colour, which implies that an uncertainty in the colour measurement could produce a considerable error in the estimation of the BC.
The coefficients of the polynomial fits and the standard deviation around the fits are presented in Table~\ref{table:bc_calibration}.

\section{Modelling}
\label{sec:modelling}

\begin{table*}
\caption{Summary of the initial conditions of the models presented in this work.}
\label{table:models}
\centering          
    \begin{tabular}{lcccccc}
    \hline\hline\noalign{\smallskip}
    Model & \mdot\ & \rcsm\ & \rcsm\ & $\beta$ & $M_{\rm CSM}$ & $t_{\rm dis}$ \\
     & [\msunyr] & [\rsun] & [cm] & & [\ms] & [days] \\
    \hline\noalign{\smallskip}
    m15\_w0.3\_r2500         & 0.3              & 2500  & 1.7\tim10$^{14}$ & 0 & 0.08 & 1.4 \\
    m15\_w0.5\_r2500         & 0.5              & 2500  & 1.7\tim10$^{14}$ & 0 & 0.14 & 1.5 \\
    m15\_w1.0\_r2500         & 1.0              & 2500  & 1.7\tim10$^{14}$ & 0 & 0.28 & 1.1 \\
    m15\_w1.0\_r3000         & 1.0              & 3000  & 2.1\tim10$^{14}$ & 0 & 0.37 & 0.7 \\
    m15\_w1.2m2\_r8000       & 1.2\tim10$^{-2}$ & 8000  & 5.6\tim10$^{14}$ & 0 & 0.02 & 6.7 \\
    m15\_w3m2\_r8000         & 3\tim$10^{-2}$   & 8000  & 5.6\tim10$^{14}$ & 0 & 0.04 & 8.5 \\
    m15\_w1.2m2\_r5000       & 1.2\tim$10^{-2}$ & 5000  & 3.5\tim10$^{14}$ & 0 & 9\tim10$^{-3}$ & 3.9 \\
    m15\_w3m3\_r12000\_beta5 & 3\tim10$^{-3}$   & 12000 & 8.4\tim10$^{14}$ & 5 & 0.23 & 6.5 \\
    m15\_w3m3\_r7000\_beta5  & 3\tim10$^{-3}$   & 7000  & 4.9\tim10$^{14}$ & 5 & 0.23 & 4.8 \\
    m15\_w1m2\_r12000\_beta5 & 1\tim10$^{-2}$   & 12000 & 8.4\tim10$^{14}$ & 5 & 0.76 & 10.0 \\
    m15\_w3m3\_r12000\_beta2 & 3\tim10$^{-3}$   & 12000 & 8.4\tim10$^{14}$ & 2 & 0.05 & 5.0 \\
    \noalign{\smallskip}
    \hline
    \end{tabular}
    \tablefoot{\mdot\ is the progenitor mass-loss rate, \rcsm\ is the extension of the wind material, $\beta$ is the wind acceleration parameter ($\beta$\eq0 corresponds to steady-state winds), $M_{\rm CSM}$ is the CSM mass, and $t_{\rm dis}$ is the theoretical epoch of disappearance of interacting lines.}
\end{table*}

The early-time bolometric light curve of SN~2023ixf allows us to constrain its progenitor mass-loss history by comparing models with observations.
Theoretical light curves are calculated using a code that solves the hydrodynamical equations assuming spherical symmetry coupled to the radiation transfer equations in the diffusion approximation \citep{bersten+11}.
The explosion is simulated by injecting energy near the centre of the progenitor star, producing a powerful shock wave that propagates out. 

In addition to bolometric light curves, our code calculates expansion velocities at different layers. Therefore, we also compare the expansion velocity at the photospheric layer to the \ion{Fe}{II}~$\lambda$5169 line velocity, given that this line gives a good estimation of the photospheric velocity \citep{dessart+05}.
The omission of the expansion velocities in the fitting procedure can result in solutions that are not consistent with the SN expansion rate \citep[see][]{martinez+20}. If this is the case, the solution found is spurious.
In order to measure expansion velocities of SN~2023ixf, we used public spectra from the WISeREP\footnote{\url{https://www.wiserep.org/}} archive \citep{yaron12} in those epochs where \ion{Fe}{II}~$\lambda$5169 profiles started appearing (approximately at 25~days after explosion).  
We used three spectra uploaded to WISeREP from the Dark Energy Spectroscopic Instrument (DESI; \citealt{levi19}) at the 4m Mayall Telescope at Kitt Peak National Observatory and one spectrum uploaded by TNS, without information about the telescope and instrument listed. We measured the expansion velocities of \ion{Fe}{II}~$\lambda$5169 in the spectra by fitting a Gaussian to the minimum of the absorption profiles. 
Additionally, we utilised the relation by \citet{faran+14a} that predicts the photospheric velocity at 50~days post-explosion from \ion{Fe}{ii} velocity measurements.

Progenitor models at the time of core collapse are needed to initialise the explosion. 
In this context, we used the public stellar evolution code \texttt{MESA}\footnote{\url{http://mesa.sourceforge.net/}} version~22.6.1 \citep{paxton+11,paxton+13,paxton+15,paxton+18,paxton+19,jermyn+23} to obtain a non-rotating RSG pre-SN model at solar metallicity \citep[Z$_{\odot}$\eq0.0142;][]{asplund+09} for a star of 15\,\ms\ on the main sequence. 
The choice of this initial mass value was carried out to agree with the progenitor luminosity observed in pre-explosion images \citep{jencson+23,neustadt+23,vandyk+23b,xiang+23}.
The stellar models were evolved from the main sequence to core collapse, defined as the time when any location inside the iron core reaches an infall velocity of 1000\,km\,s$^{-1}$. 
During massive-star evolution, mass loss was treated using the `Dutch' wind scheme defined in \texttt{MESA} \citep{vink+01,dejager+88}.
Convection was modelled using the mixing-length theory \citep{bohm58} adopting a mixing-length parameter $\alpha_{\rm mlt}$\eq2.0. The convective regions were determined using the Ledoux criterion.
Semiconvenction was implemented as a diffusive process adopting an efficiency of $\alpha_{\rm sc}$\eq1.0 \citep{langer+83}.
Convective-core overshooting is treated in the step formalism during hydrogen- and helium-core burning adopting overshooting parameters of $\alpha_{\rm os}$\eq0.15 \citep{martins+13} and 0.03 \citep{li+19} pressure scale heights, respectively. 
For later core-burning stages, we adopted the decreasing exponential approach implemented in \texttt{MESA} to account for convective overshooting with a parameter $f$\eq0.003 \citep{farmer+16,jones+17}.
The evolution of the initially 15\ms\ star with the above evolutionary parameters results in a progenitor model with a final mass of 12.7\,\ms, hydrogen-rich envelope of 8.0\,\ms, and radius of 918\,\rsun.

The observation of narrow emission lines in early-time SN spectra result from the presence of a dense and confined CSM surrounding the progenitor star. The CSM formation is thought as a consequence of a high mass-loss rate occurred during the last years to decades before core collapse, although the exact mechanism is unclear.
As the SN ejecta interacts with the CSM, kinetic energy of the ejecta is converted to radiation that can ionise the CSM and boost the SN early-time luminosity.
Given that the progenitor models computed with \texttt{MESA} do not consider the mass loss producing the CSM we are interested in, we artificially attached a CSM profile to the outer layers of the pre-SN model as usually done in the literature \citep[e.g.,][]{moriya+11,morozova+18,englert+20}.

Before attaching any CSM profile, we computed several models with different explosion energies ($E_{\rm exp}$) and compared them to our observations. Particularly, we look for agreement to the observed expansion velocities of SN~2023ixf, since these observables are strongly influenced by the energy of the explosion (for a fixed pre-SN model). We choose an explosion energy of $E_{\rm exp}$\eq1.25\tim10$^{51}$\,erg for each of our simulations with CSM interaction\footnote{We note that the inclusion of CSM can alter the photospheric velocities of a SN due to the conversion of kinetic energy into radiation at the shock front. This depends on the adopted physical parameters for the CSM.}.
We note that the ejecta mass and the explosion energy are rough estimates because in the present paper we are focused on the properties of the CSM.
In a forthcoming paper we analyse the complete evolution of the bolometric light curve and estimate the physical properties of the progenitor and explosion (Bersten et al. in prep.).

In the following we aim to reproduce the early-time bolometric light curve of SN~2023ixf by considering two different scenarios to simulate the CSM formation: steady-state (Sect.~\ref{sec:steady}) and accelerated (Sect.~\ref{sec:accelerated}) winds.
The nomenclature is based on naming each model according to its initial mass, wind mass-loss rate, radial CSM extension, and velocity law for the wind velocity.
For example, m15\_w3m3\_r12000\_beta5 corresponds to a initial mass of \mzams\eq15\,\ms, a mass-loss rate of \mdot\eq3\tim10$^{-3}$\,\msunyr, \rcsm\eq12000\,\rsun, and velocity law $\beta$\eq5.
A summary of the presented models is found in Table~\ref{table:models}.
We note that none of the models presented in this section can reproduce the change in the slope during the rise to maximum, observed in the bolometric light curve before day~2.9 post-explosion (see Sect.~\ref{sec:blc}).

We note that the explosion epoch of SN~2023ixf is based on the first detection and last non-detection, while in our models the explosion epoch is defined as the moment when the energy is deposited near the centre of the progenitor star.
Given the difference in the definition of `explosion epoch' and that it takes a few days for the shock wave to break out from the CSM, we shifted our models to match the time of maximum luminosity. These shifts were always less than 1~day for the best-fitting models.


\subsection{Steady-state winds}
\label{sec:steady}

\begin{figure}
\centering
\includegraphics[width=0.49\textwidth]{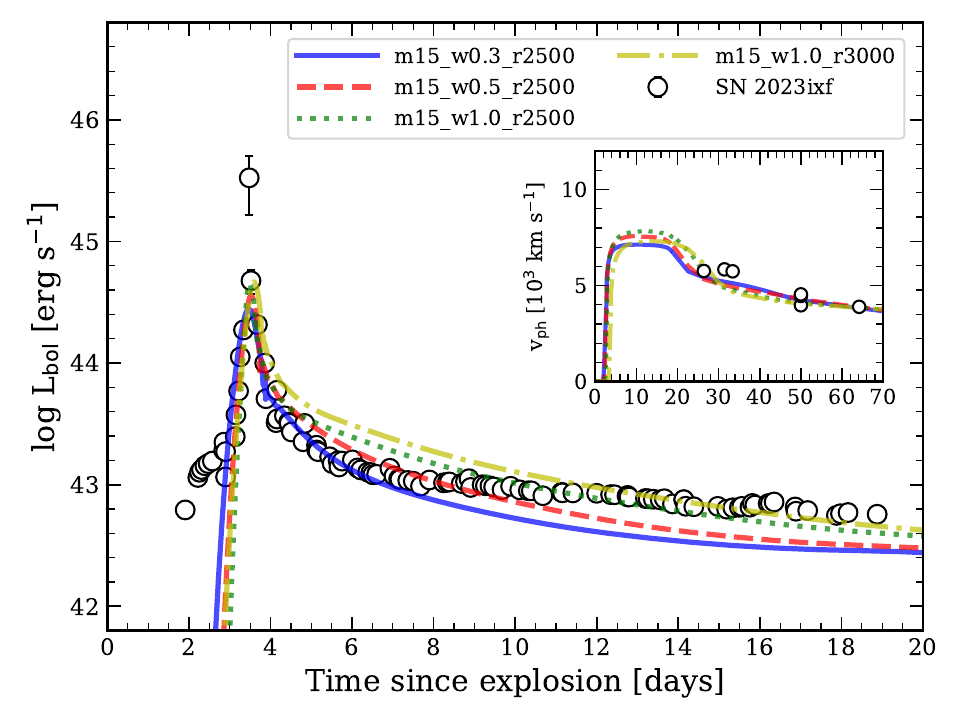}
\includegraphics[width=0.49\textwidth]{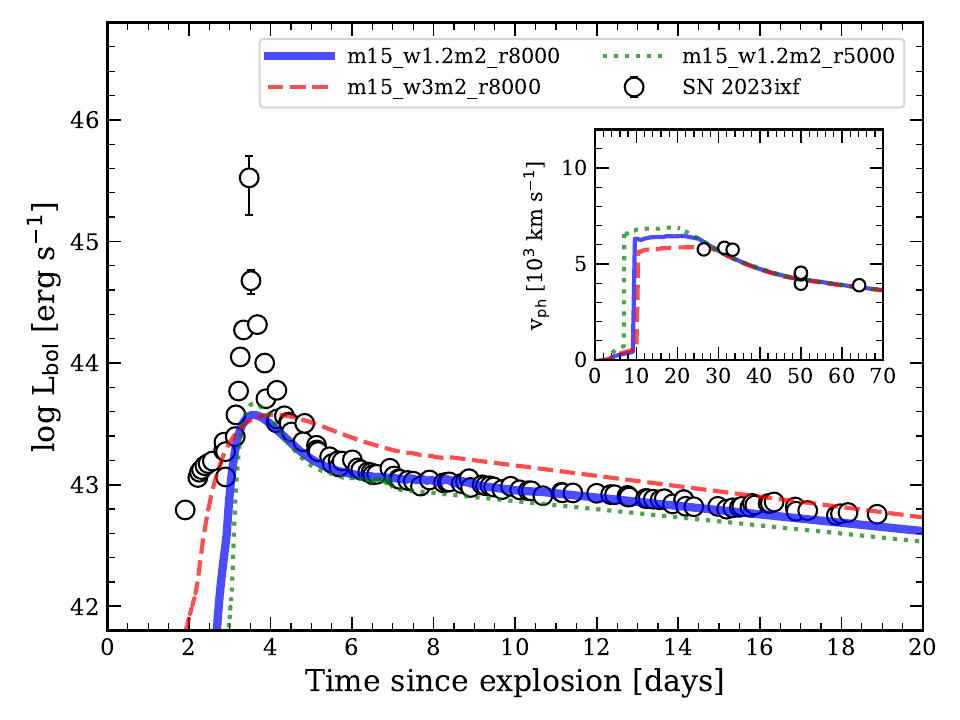}
\caption{Comparison between the bolometric light curve of SN~2023ixf (dots) with models varying the CSM properties (lines), assuming steady-state mass loss. The upper panel involves models with higher mass-loss rates and more confined CSMs than the models shown in the bottom panel. The inset plot compares the photospheric velocities of SN~2023ixf to the same models previously mentioned. For the model nomenclature, see Sect.~\ref{sec:modelling}.}
\label{fig:stat_models}
\end{figure}

The first scenario to survey involves steady-state winds. In this scenario, the CSM density ($\rho_{\rm CSM}$) is represented as $\rho_{\rm CSM}(r)$\,$=$\,$\dot{M}$/(4$\pi v_{\rm wind}r^{2}$), where $r$ is the radial coordinate, $\dot{M}$ is the wind mass-loss rate and $v_{\rm wind}$ is the velocity of the wind.
Throughout the present work, we assume a terminal wind velocity of $v_{\rm wind}$\eq115\,km\,s$^{-1}$, as measured by \citet{smith+23}.

The top panel of Fig.~\ref{fig:stat_models} compares explosion models including CSM-ejecta interaction to observations of SN~2023ixf. In this case, we choose CSM models characterised for their confined radial extent between 2500 and 3000\,\rsun\ ($\sim$1.7$-$2.1\tim10$^{14}$\,cm) and high mass-loss rates in the range of 0.3$-$1.0\,\msunyr.
While all of these models reproduce the width of the luminosity peak, model m15\_w0.3\_r2500 does it better. However, this model underestimates the luminosity after day~6 after explosion. 
Higher mass-loss rates (models m15\_w0.5\_r2500 and m15\_w1.0\_r2500) result in higher peak luminosities. However, these models achieve more luminous light curves after peak than observed, and underestimate the observed luminosities after day~10.
A more extended CSM produce higher luminosities after peak, inconsistent with observations.

The comparison from the top panel of Fig.~\ref{fig:stat_models} shows that some parts of the early light curve of SN~2023ixf can be reproduced with the adopted CSM parameters. Potentially, a more detailed study around these parameters could result in better agreements.
However, all of these models are inconsistent with the epoch of disappearance of the narrow emission lines in observed spectra.
Following \citet{dessart+17}, the narrow lines last as long as the shock is placed within a slow-moving optically-thick material (i.e., until the shock goes through the SN photosphere). 
We checked this epoch in each of our simulations and found values around $\sim$0.7$-$1.5~days after explosion, while the observations of SN~2023ixf show interacting lines until 6$-$7~days post-explosion \citep{bostroem+23}.

In the following we look for a model that reproduces the epoch when the interaction lines have faded, while matching the bolometric light curve and photospheric expansion velocities.
The thick blue solid line in the bottom panel of Fig.~\ref{fig:stat_models} shows a CSM interaction model for \mdot\eq1.2\tim10$^{-2}$\,\msunyr\ and \rcsm\eq8000\,\rsun\ ($\sim$5.5\tim10$^{14}$\,cm). This model is able to reproduce the width of the luminosity peak, post-peak luminosities, photospheric velocities, and the epoch of disappearance of the narrow emission features.
However, this model fails to reproduce the peak luminosity.
Higher mass-loss rates produce wider peaks and more luminous post-maximum light curves (bottom panel of Fig.~\ref{fig:stat_models}, dashed line). The opposite effect is expected for lower mass-loss rates.
Alternatively, a more confined CSM produce a higher peak luminosity, but lower luminosities post-maximum (bottom panel of Fig.~\ref{fig:stat_models}, dotted line).

\subsection{Accelerated winds}
\label{sec:accelerated}

\begin{figure}
\centering
\includegraphics[width=0.49\textwidth]{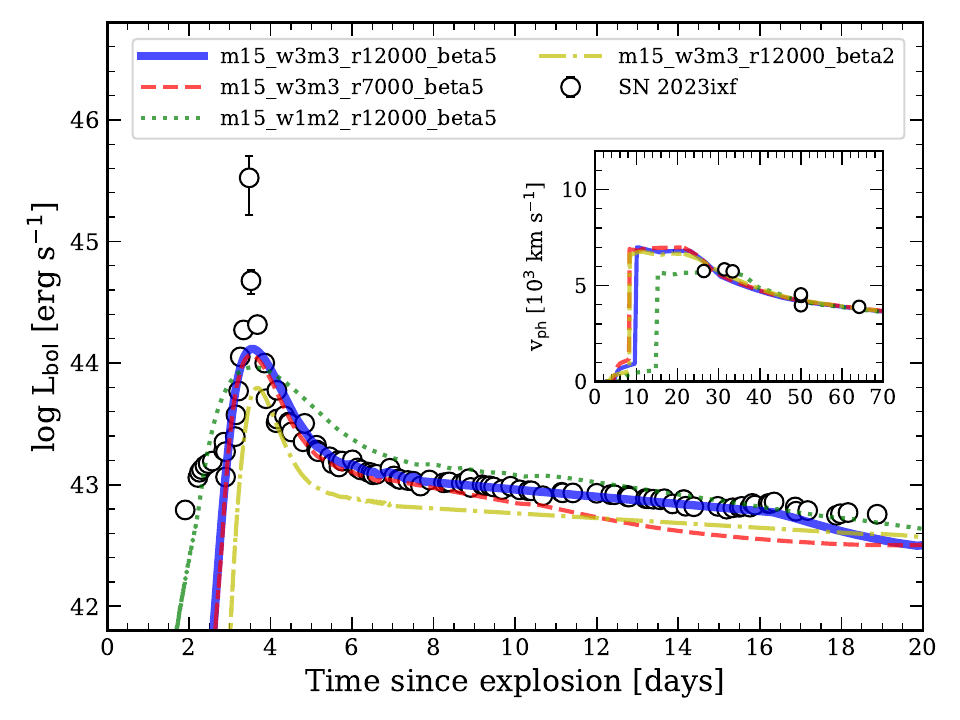}
\caption{Comparison between the bolometric light curve of SN~2023ixf (dots) with models varying CSM properties (lines) assuming wind acceleration. The inset plot compares the photospheric velocities of SN~2023ixf to the same models previously mentioned. For the model nomenclature, see Sect.~\ref{sec:modelling}.}
\label{fig:acc_models}
\end{figure}

In this section we model SN explosions within a CSM but considering the wind acceleration mechanism previously presented in \citet{moriya+18}. In this scenario, the mass-loss rate is set constant and the CSM density follows the same expression as in Sect.~\ref{sec:steady}; however, the wind velocity is no longer constant.
As in \citet{moriya+18}, the wind velocity takes the form of a $\beta$ velocity law given below:
\begin{equation}
    v_{\rm wind}(r) = v_{0} + (v_{\infty} - v_{0})\left(1 - \frac{R_{0}}{r} \right)^{\beta},
\end{equation}
where $v_{0}$ is the initial wind velocity (0.1\,km\,s$^{-1}$), $v_{\infty}$ is the terminal velocity of the wind (115\,km\,s$^{-1}$, \citealt{smith+23}), $R_{0}$ is the radial coordinate where the CSM is attached, and $\beta$ is the wind acceleration parameter \citep[see also][]{lamers_and_cassinelli99}.

We compare the early-time bolometric light curve and photospheric velocity evolution of SN~2023ixf with explosion models assuming different CSM parameters (\mdot, \rcsm, and $\beta$).
Figure~\ref{fig:acc_models} shows some of these models. The thick solid line in Fig.~\ref{fig:acc_models} represents the model m15\_w3m3\_r12000\_beta5, i.e., with \mdot\eq3\tim10$^{-3}$\,\msunyr, \rcsm\eq12000\,\rsun, and $\beta$\eq5.
From all the models we computed, this is the one that best reproduces the observations, even better than those models assuming steady mass loss (see Sect.~\ref{sec:steady}).
In addition, model m15\_w3m3\_r12000\_beta5 predicts that the narrow emission features should disappear at day~6.5, which is consistent with the observed date \citep{bostroem+23}.

Figure~\ref{fig:acc_models} also shows models computed with varying CSM properties to notice the sensitivity of the early bolometric light curve with these parameters. 
Higher mass-loss rates (e.g., model m15\_w1m2\_r12000\_beta5) produce wider peaks and more luminous post-peak light curves, while more confined CSMs (e.g., model m15\_w3m3\_r7000\_beta5) underestimate the post-peak luminosity. 
Alternatively, a smaller wind acceleration parameter (e.g., model m15\_w3m3\_r12000\_beta2) results in narrower and less luminous peak, while at the same time, less luminous light curves after maximum.
This behaviour is due to the different amount of CSM mass near the progenitor surface for varying wind acceleration parameters. A larger $\beta$ involves more mass near the progenitor surface, and therefore, a larger boost to the luminosity due to conversion of kinetic energy into radiation.
We note that the models presented cannot reach the observed maximum luminosity.

\section{Discussion}
\label{sec:discussion}

The first wind mass-loss scenario explored assumes a steady flow from the progenitor.
With the assumed wind velocity \citep[$v_{\rm wind}$\eq115\,km\,s$^{-1}$;][]{smith+23}, the size of the progenitor, and the extent of the CSM presented in Sect.~\ref{sec:modelling}, we looked for an estimate of the time before explosion in which this enhanced mass loss must have started.
For the CSM extents between 2500 and 3000\,\rsun\ ($\sim$1.7$-$2.1\tim10$^{14}$\,cm) first analysed, we found an enhanced wind that developed over the last 0.3$-$0.4\,yr before explosion\footnote{We note again that these CSM parameters do not reproduce the fading time of the narrow emission lines.}.
In addition, for a CSM extension of 8000\,\rsun\ ($\sim$5.5\tim10$^{14}$\,cm), the enhanced wind should have started $\sim$1.3\,yr before explosion.
Adopting a commonly-used wind velocity for a `superwind' (50\,km\,s$^{-1}$), the enhanced mass loss would have developed over the last $\sim$3\,yr.
The inferred timescales and mass-loss rates are similar to some values found in the literature for SN~2023ixf \citep[e.g.,][]{jacobson-galan+23,hiramatsu+23}.
However, these timescales appear to be inconsistent with pre-explosion observations.

Mid-IR \textit{Spitzer} data in the preceding $\sim$20~yr before the explosion show variability similar to those pulsating RSGs, but does not show any indication of eruptive mass-loss processes \citep{szalai+23_atel,jencson+23,kilpatrick+23,soraisam+23}.
\citet{neustadt+23} found no evidence of outbursts in optical data taken with the Large Binocular Telescope between $\sim$1 and 15~yr before the SN.
The analysis of pre-explosion optical data from the Zwicky Transient Facility \citep{bellm+19,graham+19}, Asteroid Terrestrial-impact Last Alert System \citep{tonry+18,smith+20}, Distance Less Than 40~Mpc, and All-Sky Automated Survey for Supernovae \citep{kochanek+17} surveys during the last 8~yr up to 0.3~days before explosion also found no evidence of precursor activity in the optical \citep{hiramatsu+23,dong+23,panjkov+23}.
In addition, UV observations from the \textit{Galaxy Evolution Explorer} and \textit{Swift} space telescopes did not find pre-explosion outbursts $\sim$20~yr prior to explosion \citep{flinner+23,panjkov+23}.
Therefore, pre-explosion observations indicate a quiescent progenitor in the last $\sim$20~yr, with no indication of any pre-SN outbursts or large magnitude changes ---except for the IR variability similar to pulsating RSGs \citep{soraisam+23}.
The assumption of steady-state winds results in enhanced mass loss shortly before explosion, which does not seem consistent with a quiescent progenitor.

Steady winds assume that the mass-loss rate and wind velocity are constant through the wind. However, the wind is gradually accelerated at the stellar surface until the terminal velocity is reached.
This produces an increment of the timescales for the wind development to reach a particular extension.
The bolometric light-curve modelling including CSM interaction that takes the wind acceleration into account infer that the enhanced mass loss was launched $\sim$80\,yr prior to the SN.
These timescales are related to the final stages of massive-star evolution, although the details of the connection are unknown. Some mechanisms propose mass loss driven by gravity waves excited in the convective core \citep{quataert+12,shiode+14,fuller17}, local radiation-driven instabilities in the outer layers \citep{suarez-madrigal+13}, and hydrodynamic and turbulence at pre-SN stage driven by turbulent convection \citep{smith+14}. 

\citet{teja+23} compared the $g$-band light curve of SN~2023ixf with a grid of models of \snii\ explosions interacting with an accelerated RSG wind \citep{moriya+23}.
The CSM parameters found in our study are within the ranges of values constrained by \citet{teja+23}, with the exception of the wind-acceleration parameter for which we infer a larger value.
However, \citet{teja+23} infer higher explosion energies (2$-$5\tim10$^{51}$~erg), much larger than typical \sneii\ \citep[e.g.,][]{morozova+18,martinez+20,martinez+22b} and the predictions from 1D neutrino-powered explosions \citep{sukhbold+16}. These high values could be because \citet{teja+23} did not use velocity measurements in their fitting. If expansion velocities are not taken into account in the fitting procedure, it could lead to incorrect determination of the explosion energy \citep{martinez+20}.

\citet{davies+22} carried out an analysis where they predict the characteristics of the RSGs at core collapse based on two enhanced mass-loss scenarios: a short outburst lasting a few months and a `superwind' arising from a very high mass-loss rate during the last decades prior to explosion. These authors considered an accelerated wind for the latter scenario. \citet{davies+22} found that the outburst scenario produces redder colours in a short timescale after the outburst, which would not be consistent with the steady IR variability of the progenitor of SN~2023ixf \citep[e.g.,][]{jencson+23}.
Alternatively, the scenario that involves the acceleration of the RSG winds causes redder colours decades prior the SN explosion.
\citet{jencson+23} found that the IR colours of the progenitor of SN~2023ixf are well reproduced by one of the `superwind' models from \citet{davies+22}, which assumes the same wind acceleration mechanism than the one analysed in our work. 
The discussion provided in this section would imply that the enhanced mass loss started decades before core collapse, supporting the wind acceleration scenario.

Regarding the modelling of the bolometric light curve peak, we note that none of our models can reproduce the maximum luminosity of $L_{\rm bol}$\eq3\tim10$^{45}$\,erg\,s$^{-1}$. 
While the assumption of steady-state winds produces closer values than the accelerated winds, the former is in contradiction with the epoch of disappearance of the interacting lines as described in Sect.~\ref{sec:steady}.
This indicates that we still need to improve our knowledge of the conditions under which the CSM was formed.
It is also important to mention that some physics are not included in our code, as non-local thermodynamic equilibrium or multi-dimensional effects \citep[see][for detailes]{bersten+11}.

\section{Summary and conclusions}
\label{sec:conclusions}

SN~2023ixf is among the closest \snii\ in the last decades, which allowed intensive multi-wavelength and high-cadence observations.
We used publicly available data to calculate the early ($<$19\,days post-explosion) bolometric light  curve based on the integration of the observed SED (from UV to NIR bands) and black-body extrapolations for the unobserved flux at shorter and longer wavelengths.
Thanks to the early monitoring and high cadence of observations, we capture the sudden rise to maximum and the successive fall of the bolometric light curve. This is the first time that this behaviour is observed in bolometric luminosities given the lack of early-time multi-wavelength observations for most \sneii.

The fact that there are a small number of \sneii\ with detailed calculations of their early bolometric light curve (see e.g., \citealt{yaron+17} and \citealt{jacobson-galan+22} for bolometric light curves after maximum for SN~2013fs and SN~2020tlf, respectively), allowed us to test the currently available calibrations of BC against colours.
This analysis provides good agreements for most of these calibrations.
Additionally, we included the observations of SN~2023ixf to the recently-published calibrations of BC from \citet{martinez+22a}. These calibrations include data of 74~\sneii, but none with observations as early as SN~2023ixf.
Therefore, the incorporation of SN~2023ixf to the previously-mentioned calibrations allows to extend them to bluer optical colours, and therefore, to earlier epochs.
It would be necessary to include all \sneii\ with early detections and good photometric coverage in order to analyse the bluest part of these calibrations in detail, and to study a possible general behaviour.

Armed with the bolometric light curve for SN~2023ixf, we have studied the mass-loss history of the progenitor of SN~2023ixf through comparison with hydrodynamical simulations of \snii~explosions with CSM interaction.
We found that a CSM interaction model that takes the wind acceleration into account with \mdot\eq3\tim10$^{-3}$\,\msunyr, \rcsm\eq12000\,\rsun, and $\beta$\eq5 reproduces the width of the luminosity peak, the post-peak luminosity, and the epoch of disappearance of the interaction lines in the spectra.
Our findings indicate an enhanced wind that developed continuously over the last $\sim$80~yr of the progenitor evolution.
This may be consistent with the quiescent of SN~2023ixf in the last 20~yr prior to explosion, favouring the accelerated wind scenario ---in connection with the results of \citet{jencson+23}.
In a forthcoming paper, we analyse the complete bolometric light curve and photospheric velocity evolution of SN~2023ixf and derive the physical properties of the progenitor and explosion, which allow us to have a full description of the nature of SN~2023ixf (Bersten et al. in prep.).

\begin{acknowledgements}
L.M. acknowledges support from a CONICET fellowship.
L.M. and M.O. acknowledges support from UNRN~PI2022~40B1039 grant.

This research has made use of the Spanish Virtual Observatory (\url{https://svo.cab.inta-csic.es}) project funded by MCIN/AEI/10.13039/501100011033/ through grant PID2020-112949GB-I00.
This work has made use of WISeREP (\url{https://www.wiserep.org}).
\\
\textit{Software:} \texttt{NumPy} \citep{numpyguide2006,numpy2011}, \texttt{matplotlib} \citep{matplotlib}, \texttt{Astropy} \citep{astropy2013,astropy2018,astropy2022}, \texttt{SciPy} \citep{scipy}, \texttt{emcee} \citep{emcee}, \texttt{Pandas} \citep{pandas}, \texttt{ipython/jupyter} \citep{jupyter}, \texttt{extinction} (\url{https://github.com/kbarbary/extinction}).
\end{acknowledgements}

\bibliographystyle{aa.bst}
\bibliography{biblio.bib}

\end{document}